# BORON-DOPED GRAPHENE AS ACTIVE ELECTROCATALYST FOR OXYGEN REDUCTION REACTION AT A FUEL-CELL CATHODE

Gianluca Fazio, Lara Ferrighi, Cristiana Di Valentin[*]

*Dipartimento di Scienza dei Materiali, Università di Milano-Bicocca*

*via Cozzi 55 20125 Milano Italy*

**Abstract**

Boron-doped graphene was reported to be the best non-metal doped graphene electrocatalyst for the oxygen reduction reaction (ORR) working at an onset potential of 0.035 V [JACS 136 (2014) 4394]. In the present DFT study, intermediates and transition structures along the possible reaction pathways are determined. Both Langmuir-Hinschelwood and Eley-Rideal mechanisms are discussed. Molecular oxygen binds the positively charged B atom and forms an open shell end-on dioxygen intermediate. The associative path is favoured with respect to the dissociative one. The free energy diagrams along the four-reduction steps are investigated with the methodology by Nørskov and co. [JPC B 108 (2004) 17886] in both acidic and alkaline conditions. The pH effect on the stability of the intermediates of reduction is analyzed in terms of the Pourbaix diagram. At pH = 14 we compute an onset potential value for the electrochemical ORR of $U$ = 0.05 V, which compares very well with the experimental value in alkaline conditions.

**Graphical abstract**

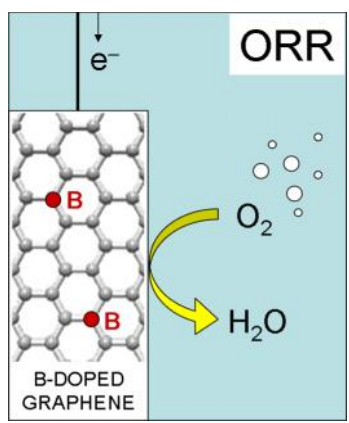

---

[*] Corresponding author: cristiana.divalentin@mater.unimib.it, +390264485235.

# 1. Introduction

Platinum (Pt) and its alloys are still the most widely used electrocatalysts for the oxygen reduction reaction (ORR) in fuel cells and metal-air batteries. However, the major drawbacks to their use, the high-cost, scarcity and large overpotential, are the main reasons for the intensive efforts devoted by the scientific community in finding efficient, metal-free and cheap electrocatalysts for ORR and other redox reactions [1]. Doped-graphene systems have been recently proposed in the literature as promising alternatives [2,3,4,5,6,7,8,9], being potentially active catalysts for ORR. Among them, N-doped graphene has attracted most of the attention and research work [2,4,6,7,8], although, more recently, B-doped graphene systems have also been reported to efficiently catalyzed this reaction [10,11]. Some comparative experimental data [12] have just appeared in the literature showing that B-doped graphene is even slightly better than N-doped one, both definitely surpassing the performance of other non-metal (O-, P-, S-) doped graphene systems. The details of the reaction path are yet not very clear [13,14]. Moreover, they could be different for different doping elements. In the case of N-doped graphene [15,16,17,18,19,20,21,22], the dissociative path is generally considered not to be a viable route to products, even though no clear-cut evidence definitely proves that. As regards the associative mechanism, the more efficient direct 4 $e^-$ pathway is found to compete with the 2 $e^-$ one.

The molecular oxygen adsorption step, which is the first step for both dissociative and associative mechanisms, is considered to be critical in the ORR process. On pure graphene (G), it goes through the formation of an intermediate side-on bridging dioxygen species (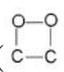) paying a very high energy cost (about +2.5 eV with respect to G + $O_{2(g)}$) [21,23] which is essentially the reason for the total inertness of graphene towards molecular oxygen. In the case of N-doped graphene (NG), which is the most studied and known case of doped graphene, the analogous species (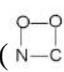) is still rather high in energy (about +1.6 eV with respect to NG + $O_{2(g)}$) [21]. In most theoretical works [17,19,20], this preliminary and very energetically expensive step is overlooked or neglected, in favour of the much more stable hydroperoxo species (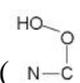). This oversimplification of the reaction path is a serious issue, primarily because it leads to incorrect conclusions. Some authors [21,22], recognizing the limitations of bulk graphene N species as attractive sites for $O_2$ adsorption, have proposed a prominent role of the quaternary N atoms at the graphene edges, especially in the case of N-doped graphene nanoribbons (NR). They find that at the NR edges a stable end-on dioxygen intermediate (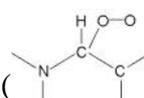) can be formed with a much lower

adsorption energy barrier (only 0.35 eV) than the side-on one in the NG bulk (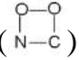). The end-on species was proposed to be the first intermediate for $O_2$ adsorption in both dissociative and associative paths for ORR.

Other critical issues in the use of doped graphene as electrocatalysts for ORR regard i) the activation barriers for the proton/electron transfers along the full reduction process, arising at the equilibrium potential, which require a potential bias to be applied to the fuel cell and ii) the selectivity towards the 4 $e^-$ associative path. These issues can be addressed and analyzed by applying a methodology which has been first developed by Nørskov and co. [24] for ORR on the paramount case of Pt electrodes. This method in combination with density functional calculations provides a description of the free-energy landscape of the electrochemical ORR over the electrocatalyst surface as a function of the applied bias and suggests ways to improve the electrocatalytic properties of fuel-cell cathods.

In this work we present a detailed study of the possible reaction paths for ORR as catalyzed by B-doped graphene, through the identification of all the intermediates and transition structures for both dissociative and associative mechanisms. Additionally, the methodology by Nørskov and co. [24], which is generally applied to metal electrodes, is presently used to analyzed the B-doped graphene catalyst. This approach allows to determine the critical reduction steps along the overall process and to make a comparison with the analogous quantities for Pt electrodes. Finally, we also present the Pourbaix diagram [25] to identify the dependence of the potential for each step of reduction [4 × ($H^+$ + $e^-$)] with the pH value of the electrolytic solution.

We find that the $O_2$ adsorption on B-doped graphene goes through an end-on dioxygen intermediate, never observed for both G and NG, involving the bulk graphene B atom (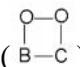), which is preliminary to the formation of the high energy side-on (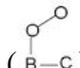) intermediate for the dissociative mechanism and of the hydroperoxo (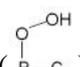) species for the associative one. The existence of this end-on adsorbed $O_2$ species is crucial to trigger the associative ORR path which we consider to be the reason for the excellent performance of B-doped graphene as an electrocatalyst for this reaction [10,11,12].

## 2. Computational details

All the calculations were performed with the GAUSSIAN09 [26] (G09) suite of programmes and the B3LYP [27,28] functional. Spin polarization is taken into account in the case of open shell systems. The model for pure graphene is a circumcoronene molecule ($C_{54}H_{18}$). In a previous work we have shown through a detailed comparison of molecular calculations with periodic systems that the energetics of oxygen reactivity with graphene and B-doped graphene obtained with the different approaches is fully consistent [23]. The orbitals were described with gaussian basis functions [6-311+G* for the inner five C atoms, the B atom, and the O and H atoms involved in the water formation; 6-31G* for the rest of the model]. All atoms were allowed to relax during the geometry optimization without any symmetry constraint.

Critical points on the potential energy surface have been characterized by diagonalizing the Hessian matrices. The transition-state structures were searched by numerically estimating the matrix of the second-order energy derivatives at every optimization step and by requiring only one eigenvalue of this matrix to be negative.

Vibrational frequencies in the harmonic approximation were calculated for all optimized structures and used, unscaled, to compute zero point energies, enthalpies and Gibbs free energies.

The reference electrode is the standard hydrogen electrode SHE [24]. At pH = 0 and at potential $U = 0$ V vs SHE, the reaction $H^+ + e^- \leftrightarrow \frac{1}{2} H_2$ is in equilibrium at 1 bar $H_{2(g)}$ at 298 K, thus G ($H^+ + e^-$) = G ($\frac{1}{2}H_2$) in these conditions. The free-energy difference of the full ORR with the present setup is -4.64 eV which we consider in more than satisfactory agreement with the experimental value of -4.92 eV. The free energy of $OH^-$ is derived as G ($OH^-$) = G ($H_2O_{(l)}$) - G ($H^+$), where G ($H^+$) is corrected by - $kT \times \ln 10 \times$ pH, to account for the pH conditions.

The contribution of bulk solvent (water) effects to the Gibbs free energy ($G_{sol}$) was computed using the polarizable continuum model (PCM) in the SMD version [29,30] implemented in the Gaussian09 package. Small structural modification are observed as a consequence of the relaxation in the solvent. The effect of dispersion forces, estimated with the Grimme D3 method [31,32] as implemented in the Gaussian09 code, was found to be analogous for the intermediates of the ORR and therefore negligible on the $\Delta G$ values.

## 3. Results and Discussion

*3.1 Oxygen Reduction Reaction Paths*

The pathways for the oxygen reduction reaction may be a) dissociative or b) associative.

The dissociative pathway is generally considered to go through the following steps (* = surface):

$$O_2 + * \rightarrow *O_2 \qquad (1)$$

$$*O_2 \rightarrow *O + *O \qquad (2)$$

$$*O + *O + H^+ + e^- \rightarrow *O + *OH \qquad (3)$$

$$*O + *OH + H^+ + e^- \rightarrow *O + H_2O \qquad (4)$$

$$*O + H_2O + H^+ + e^- \rightarrow *OH + H_2O \qquad (5)$$

$$*OH + H_2O + H^+ + e^- \rightarrow 2H_2O + * \qquad (6)$$

We have investigated this reaction path on the BG surface as described in details in Figures 1 (eq. 1-4) and 3 (eq. 5-6), where the free energy variation ($\Delta G_{sol}$) of the various intermediates and transition structures is reported along the reaction coordinate, with the reference free energy set at the value of the starting reagents, i.e. $G_{sol}$ [BG + $O_2$ (g) + $4H^+$ + $4e^-$]. Molecular oxygen in its triplet state bind to the positively charged B atom (according to NBO charges: 0.36) on the BG surface, forming an end-on dioxygen species in a doublet spin configuration state ( 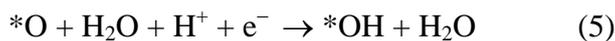 which will be called BGOO or species **2**), without any spin crossing issues. The residual spin density is localized entirely on the two oxygen atoms (see Figure S1 in the Supplementary material). This intermediate can evolve to a less stable side-on dioxygen species ( 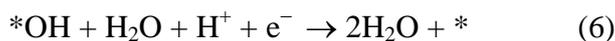, **3**) through a transition state (**2TS**) with a 1.04 eV activation free energy. This step is required in order to proceed to the O-O bond breaking (eq. (2)), which is the most energetically expensive (rate determining) step, with an activation free energy of 1.13 eV, to form a very stable oxidized boron species (-C-O-B-O-C-, **4**). Then, the first reduction step takes place (+ $H^+$ + $e^-$), again through a rather high free energy barrier (**5TS**) to form the hydroxyl species (-C-O-B-OH-C-, **6**). Note that, as starting point for this reduction step (and also for the next ones), we now consider the proton as adsorbed on a reduced surface (intermediates **5**, **7**, **12**, **14**), according to a Langmuir-Hinschelwood (LH) mechanism. In the next paragraphs we will also discuss the Eley-Rideal (ER) mechanism where the proton comes directly from the electrolyte. The second reduction step (+ $H^+$ + $e^-$) is much easier (with 0.99 eV barrier, **7TS**) and leads to a very stable BGO or **8** species (-C-O-B-C-) and one $H_2O$ molecule. As described in Figure 3, from BGO, through other two reduction steps with a first rather high free energy barrier (1.13 eV, **12TS**) and a second lower one (0.63 eV, **14TS**), the reaction proceeds to BGOH (**13**) first and then to fully reduced BG and a second $H_2O$ molecule. The overall free energy variation for the complete ORR is -4.64 eV, as discussed in the Computational details. Note that various possible adsorption sites for the additional H atom in the two reduction steps (intermediates **5** and **12**) have been compared and the configurations reported with the H atom on the C in ortho position to B are the most stable. It is very important to stress that all the intermediates (**2**, **4**, **6**, **8**, **9**, **13**) which are expected to accept the extra electron (and then the $H^+$) have a very positive electron affinity (EA > 2.0 eV) defined as the energy difference between the neutral species and the singly negatively charged one (e.g. E (**2**) - E (**2$^>$**) = EA (**2**), see Table 3).

The associative pathway can be either a 2 e⁻ or 4 e⁻ reduction process. The complete reduction requires the following steps (* = surface):

$$O_{2(g)} + * \rightarrow *O_2 \qquad (7)$$

$$*O_2 + H^+ + e^- \rightarrow *OOH \qquad (8)$$

$$*OOH + H^+ + e^- \rightarrow *O + H_2O \qquad (9)$$

$$*O + H_2O + H^+ + e^- \rightarrow *OH + H_2O \qquad (10)$$

$$*OH + H_2O + H^+ + e^- \rightarrow 2H_2O_{(l)} + * \qquad (11)$$

The first adsorption step (1) is analogous to the dissociative mechanism and leads to the formation of the end-on dioxygen species (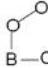, BGOO, **2**), see Figures 2. As a next step, this species is directly reduced (+ H⁺ + e⁻) to a hydroperoxo species (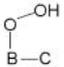, BGOOH, **9**), without the need to go through the very energetically expensive side-on intermediate as in the dissociative mechanism (Figure 1). This is essentially the reason why the associative mechanism can be safely considered the dominant one for the ORR process on B-doped graphene. Analogous intermediates with the OOH binding a C atom in ortho and para position with respect to the B atom have been considered, but were found to be at least 0.7 eV higher in energy (see Supplementary Material). As described in Figure 2, starting from the very stable BGOOH intermediate (**9**) we have investigated various possible paths for the second step of reduction (+ H⁺ + e⁻), with the additional H atom adsorbing on the C in ortho position with respect to B (the most stable adsorption site resulting in species **11**, see Table S1 in the Supplementary material). Excluding the direct dissociation to form -C-O-B-COH- species (**10**) due to a too high free energy barrier (2.02 eV), the two feasible paths are: (a) the formation of $H_2O_2$ (also referred to as "$H_2O_2$ *mechanism*"), leaving the restored BG catalyst, through a relatively low activation free energy of 0.76 eV (**11TS₂**) and an overall thermodynamic energy release of -1.42 eV; (b) the formation of a first $H_2O$ molecule, leaving an oxidized catalyst (BGO, **8**), through a transition structure (**11TS₁**) with an activation free energy of 0.68 eV but an overall thermodynamic energy gain of -2.98 eV. The first, (a), is a 2 e⁻ pathway. The released $H_2O_2$ can further dissociate, by overcoming a free energy barrier of 0.92 eV, into two adsorbed OH species (**15**, Figure 2) which could eventually be involved in the reaction described in eq. (11) from the analogous intermediate **13** going to products (Figure 3). In the second case, (b), the reaction can proceed through the next two reduction steps (4 e⁻ pathway) as in equations (10)-(11) which are totally analogous to equations (5)-(6) for the dissociative mechanism and have been already discussed above and represented in Figure 3. As already mentioned, however, the third step of reduction is characterized by a very high activation barrier which is expected to be the bottleneck for the 4 e⁻ pathway, stopping the reduction process to the first two reduction steps and leaving a very stable BGO species.

In order to emphasize the important role played by the solvent, in Tables 1 and 2 we report both the free energies of reaction and free activation energies in gas-phase ($\Delta G$ and $\Delta G^{\ddagger}$) and after correction with the solvent effect ($\Delta G_{sol}$ and $\Delta G^{\ddagger}_{sol}$) for some reaction steps of the pathways in Figures 1, 2 and 3. We observe that the inclusion of the dielectric stabilizes BGOO (**2**), BGOOH (**9**) and BGOH (**13**) species. Some of the activation barriers result to be unchanged, while that for **5TS** is reduced by 0.25 eV, and those for **7TS, 11TS$_2$** and **14TS** are enhanced as a consequence of the large stabilization of BGOH by the solvent environment. This analysis shows that gas-phase calculations, which are often used to interpret experiments performed in solution, may not give the correct and accurate picture of the reaction paths.

The hypothesis that the reaction proceeds through a LH mechanism is supported by the fact that there is a considerable energy gain for a H$^+$ ion to be adsorbed on the reduced surface with respect of being solvated in bulk water as H$^+_{(aq)}$ (estimated to be about -1.3 eV with respect to the proton solvation free energy: G(H$^+$)$_{aq}$ = -11.72 eV, as derived from experiments [33,34]). However, since very high activation barriers are computed along the LH reaction path, the ER mechanism, where the proton comes from the electrolyte, becomes a reasonable alternative. The activation barrier for proton transfers is expected to be low, but its computation is very demanding since it requires an accurate atomistic description of bulk water. Jacob and co. [35] have estimated this barrier to be about 0.3 eV with a CCSD(T) calculations. In the present work we will approximate all protonation steps barriers along the ER mechanism to be 0.3 eV (see dashed red curves in Figures 1, 2 and 3). Since Jacob and co. have studied the ORR on the Pt surface with the same functional (B3LYP) as in the present work, we report a direct comparison for the energy of the various intermediates in the two cases, with respect to the ideal situation where no overpotential is expected (Figure S2 in the Supplementary Material). This analysis suggests that BG is closer to the ideal ORR electrocatalyst than Pt.

Since for molecular calculations it is possible to compute charged systems, we have further investigated the electron/proton transfers by decomposing the reduction process in two parts: first we simulate the electron transfer (from an ideal hydrogen electrode, SHE) by producing a negatively charged species; secondly, we describe the proton transfer by adding a H$^+$ (coming from the aqueous solution) to the system. For the associative mechanism, this leads to a new reaction scheme where the charged intermediates are shown at each reduction step (Figure 4). The neutral intermediates preceding or following the electron/proton transfers are the same as those in Figures 2 and 3. We find that the electron transfer from the standard electrode potential costs about 0.3 eV in the first, second and fourth reduction steps. For the third step we find that BGO is isoenergetic with BGO⁻, with no cost for the electron transfer. These results nicely correlate with the electron affinities of the species involved which are reported in Table 3.

*3.2 Free Energy Diagrams at varying Electrode Potential*

In the following we will analyze the ORR process on B-doped graphene by using the methodology developed by Nørskov and co. [24] for metal surface catalysts. For this procedure we have used the same conditions to evaluate the free energies for the intermediates of the electrochemical reactions, as described in the Computational details. Here, however, the free energy reference level is set at the value of the reaction products (G[BG + 2H$_2$O] for acidic conditions and G[BG + 4 OH⁻] for alkaline conditions) and some additional effects are taken into account. 1) In order to model the effect of the water environment, we have introduced six explicit water molecules which approximately simulate a water monolayer over the surface and allow the establishment of direct hydrogen bonding with the oxygenated surface species ($\Delta G_w$). Additionally, the effect of the bulk water has been also described with the PCM-SMD method ($\Delta G_{w/sol}$).

2) The bias effect on the energy of all states involving an electron in the electrode is included by shifting the energy of the state by >$neU$, where $U$ is the electrode applied potential relative to standard hydrogen electrode (SHE), $e$ is the elementary charge and $n$ is the number of proton-electron pairs transferred. In acidic conditions, values of $U$ range from 0 V, corresponding to the reaction running by short circuiting the cell, to 1.23 V, corresponding to the situation where the fuel cell has the maximum potential allowed by thermodynamics or equilibrium potential $U_0$. In basic conditions (pH = 14), values of $U$ range from 0 V to 0.40 V. Since some of the steps are uphill at equilibrium potential, an applied potential $U$ is required to overcome the positive free energy change, determining an overpotential ($\eta = U_0 - U$).

3) The effect of a pH different from 0 of the electrolytic solution is taken into account by correcting the free energy of H⁺ ions for the concentration dependence of the entropy: G(pH) = -kT × ln 10 × pH.

According to the methodology by Nørskov and co. [24], the free energy diagram is constructed by taking into account the starting reagents, the various intermediates at each stage of reduction and final products of reaction (for the ball and stick models of the intermediates see Figure 5). We have considered the overall ORR process in both acidic and alkaline conditions, as reported in Figure 6 and Figure 7.

In acidic conditions the process is represented by the following equation:

$$O_{2(g)} + 4\ (H^+ + e^-) \rightarrow 2\ H_2O_{(aq)} \qquad (12)$$

The corresponding free energy diagram is shown in Figure 6 and the thermochemistry data is presented in Table 4. We observe that at $U = 0$ V, each step of reduction is downhill in terms of $\Delta G_{w/sol}$. Note that there is an energy cost (0.46 eV) to adsorb $O_2$, preceding the first reduction step, which, however, is not dependent on the applied potential. With respect to an ideal catalyst (represented by a dashed line in Figures 6), BGO and BGOH are a bit too low in energy, while BGOOH is slightly too high. At the equilibrium potential of $U_0 = 1.16$ V (exp. 1.23 V), some intermediates have negative $\Delta G_{w/sol}$, thus some steps become uphill, as mentioned above. We observe that the lowest working potential, or onset potential, for a downhill reduction process is $U^{onset} = 0.87$ V, corresponding to an overpotential of $\eta = 0.29$ V. It is interesting to analyze the variation in free energy differences along the path as a consequence of the introduction of the water environment, as summarized in Table 4. The effect of explicit water molecules is clearly that of stabilizing the BGOOH and BGOH species with respect to BGOO and BGO.

In alkaline conditions (pH = 14) the ORR process is described by the following equation:

$$O_{2(g)} + 2\ H_2O_{(aq)} + 4\ e^- \rightarrow 4\ OH^- \qquad (13)$$

The corresponding free energy diagram is shown in Figure 7 and the thermochemistry data is presented in Table 4. We observe that at $U = 0$ all the reduction steps are downhill in terms of $\Delta G_{w/sol}$. BGOO and BGOOH are too high in energy than what one would desire, while BGO is a bit too stable as well as BGOH. Also in the case of alkaline ambient, the effect of water interaction with the chemical species is to alter their relative stability (see Table 4). Again BGOOH and BGOH result to be particularly stabilized with respect to BGOO and BGO. At the equilibrium potential, $U_0$, three steps become uphill with positive $\Delta G_{w/sol}$ (the first reduction step is the most uphill by +0.28 eV). The onset potential for a downhill reduction process is found to be $U^{onset} = 0.05$ V, corresponding to an overpotential of $\eta = 0.28$ V. This value perfectly matches the experimental measured onset potential ($U^{onset} = 0.035$ eV) recently reported in the literature for ORR on a B-doped graphene electrocatalyst in alkaline conditions [12].

*3.3 Pourbaix Diagrams*

In this last section we present and discuss the surface Pourbaix diagram [25] which allows to analyze the free energy dependence of the various electrochemical intermediates at varying potential and pH conditions:

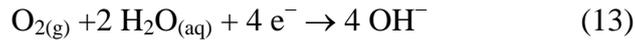
$$\Delta G_{w/sol} = \Delta G_{w/sol[U\,=\,0,\,pH\,=\,0]} - neU - kT \ln 10\ pH \qquad (14)$$

The intermediate with the lowest free energy at a given set of potential and pH determines the stable form of the surface in those conditions. Instead of plotting three-dimensional graphs, one generally first investigates the dependence with the potential fixing pH = 0, as in the top panel of Figure 8.

Each intermediate species along the catalytic path for ORR on B-doped graphene is characterized by a straight line with a negative slope which is determined by the number $n$ of electrons to be transferred (1 for BGOH, 2 for BGO, 3 for BGOOH and 4 for BGOO). In different potential ranges, a different intermediate is found to be the most stable (i.e. the lowest line at a given potential). These lowest lines cross at three different potentials: at low values of $U$ ($0 < U < 0.94$ V) the most stable form is the pure catalyst BG, in the interval $0.94$ V $< U < 1.10$ V it is BGOH, for $1.10 < U <$ 1.53 it is BGO. Note that BGOOH is never the most stable species at any potential. Values of $U >$ 1.23 V are above the equilibrium potential for the ORR which is the maximum potential allowed at pH = 0 and are, thus, not feasible.

The pH range from 0 to 14 is then investigated, considering that:

$$U = U_{pH=0} - kT \ln 10 \, pH \qquad (15)$$

The resulting surface Pourbaix diagram at T = 298 K (bottom panel of Figure 8) shows clearly which is the most stable form as a function of the pH and of the potential. The dotted line indicates the values of feasible potentials at a given pH for the ORR. At pH = 0, the BG catalyst is stable only up to a potential of $U = 0.94$ V, in line with what discussed in the previous section. At pH = 14 the catalyst BG is the most stable species only up to a potential of $U = 0.11$ V.

We have investigated the possibility that the basic species (BGO$^-$ and BGOO$^-$) could be stable at alkaline conditions by estimating the pK$_a$ of BGOH and BGOOH. The computed values are very high (> 14), as it is generally the case for alcohol and hydroperoxide species, therefore BGO$^-$ and BGOO$^-$ do not appear in the Pourbaix diagram in the pH range presented. This fact means that all reduction steps are coupled proton/electron transfers at any pH value.

## 4. Conclusions

In summary, through a hybrid density functional study it was possible to unravel the details of the reaction pathways for the ORR catalyzed by B-doped graphene. We considered both the associative and dissociative mechanisms in the LH or ER schemes. Among all the intermediates and transition structures which have been characterized along the various reaction coordinates, the most significant species is the end-on dioxygen species (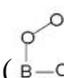), resulting from the adsorption of molecular oxygen on the positively charged B doped atom. This is an open shell intermediate which cannot be conceived in the case of pure graphene, and was never found for N-doped bulk graphene, the latter being the most competitive alternative to B-doped as an ORR electrocatalyst [12].

Starting from these data we have used the methodology developed by Nørskov and co. to determine the free energy diagrams at different electrode potentials. We concluded that in acidic conditions (pH = 0) an overpotential of 0.29 V is required to have an overall downhill reduction process. In alkaline solutions (pH = 14) the analogous overpotential is 0.28 V which corresponds to an onset potential of $U^{onset}$ = 0.05 V, in excellent agreement with the experimental reported value [12] for ORR on B-doped graphene samples. The pH effect on the electrochemistry of ORR has been also investigated in terms of Pourbaix diagrams. The analysis has clearly evidenced that in the range 0 < pH < 14: i) the BG catalyst can always be restored by applying a sufficient overpotential ; ii) BGOOH is never the most stable species at any potential iii) all the reduction steps are coupled proton/electron transfers since the deprotonated species $BGO^-$ and $BGOO^-$ are not stable.

**Appendix A. Supplementary material**

Supplementary data associated with this article can be found, in
the online version, at ....

**Acknowledgments**

We thank Lorenzo Ferraro for his technical help. This work was supported by the Italian MIUR through the national grant Futuro in ricerca 2012 RBFR128BEC "Beyond graphene: tailored C-layers for novel catalytic materials and green chemistry" and by CINECA supercomputing center though the LISA grant 2014 "LI03p_CBC4FC".

**Table 1** Thermochemistry of the reaction steps along the dissociative pathway described in Figure 1 and Figure 3. Energy and free energy variations in eV.

| DISSOCIATIVE PATHWAY | | | | | | | | |
|---|---|---|---|---|---|---|---|---|
| STEP | | | E | G | $G_{sol}$ | $E^{\ddagger}$ | $G^{\ddagger}$ | $G^{\ddagger}_{sol}$ |
| $O_{2(g)}$ + *  →  $*O_2$ | [1 | 2] | 0.34 | 0.86 | 0.54 | 0.35 | - | - |
| $*O_2$  →  *O-O | [2 | 3] | 0.77 | 0.84 | 0.60 | 0.88 | 0.93 | 1.04 |
| *O-O  →  *O + *O | [3 | 4] | -2.08 | -2.09 | -1.69 | 0.84 | 1.15 | 1.13 |
| *O + *O + *H  →  *O + *OH | [5 | 6] | -1.34 | -1.41 | -1.73 | 1.72 | 1.60 | 1.35 |
| *O + *OH + *H  →  *O + $H_2O_{(l)}$ | [7 | 8] | -0.56 | -1.55 | -1.11 | 0.77 | 0.67 | 0.99 |
| *O + *H  →  *OH | [12 | 13] | -1.00 | -1.05 | -1.36 | 1.29 | 1.16 | 1.13 |
| *OH + *H  →  * + $H_2O_{(l)}$ | [14 | 1] | -1.17 | -2.16 | -1.73 | 0.54 | 0.42 | 0.63 |

**Table 2** Thermochemistry of the reaction steps along the associative pathway described in Figure 2 and Figure 3. Energy and free energy variations in eV.

| ASSOCIATIVE PATHWAY | | | | | | | | |
|---|---|---|---|---|---|---|---|---|
| STEP | | | E | G | $G_{sol}$ | $E^{\ddagger}$ | $G^{\ddagger}$ | $G^{\ddagger}_{sol}$ |
| $O_{2(g)}$ + *  →  $*O_2$ | [1 | 2] | 0.34 | 0.86 | 0.54 | 0.35 | - | - |
| *OOH  →  *O + *OH | [9 | 10] | -0.94 | -0.85 | -0.63 | 1.91 | 1.92 | 2.02 |
| *OOH + *H  →  * + $H_2O_{2(aq)}$ | [11 | 1] | -0.95 | -1.95 | -1.42 | 0.71 | 0.56 | 0.76[a] |
| *OOH + *H  →  $H_2O_{(l)}$ + *O | [11 | 8] | -2.46 | -3.39 | -2.98 | 0.67 | 0.61 | 0.67 |
| * + $H_2O_{2(aq)}$  →  *OH + *OH | [1 | 15] | -1.25 | -0.15 | -0.71 | 0.37 | 1.36 | 0.92 |
| *O + *H  →  *OH | [12 | 13] | -1.00 | -1.05 | -1.36 | 1.29 | 1.16 | 1.13 |
| *OH + *H  →  * + $H_2O_{(l)}$ | [14 | 1] | -1.17 | -2.16 | -1.73 | 0.54 | 0.42 | 0.63 |

[a] This barrier was obtained by performing a single point calculation for the solvent effect on the gas-phase optimized transition structure.

**Table 3** Electron affinities (in eV) defined as: EA = E (SPECIES) - E (SPECIES$^>$). Both electronic energies (vacuum) and free energies corrected for the solvent (solution) are used.

|  |  | ELECTRON AFFINITIES | |
| --- | --- | --- | --- |
|  | SPECIES | VACUUM | SOLUTION |
| BGOO | **2** | 2.00 | 3.94 |
| BGOOH | **9** | 2.46 | 3.90 |
| BGO | **8** | 2.80 | 4.30 |
| BGOH | **13** | 2.39 | 3.84 |
| BGO_O | **4** | 2.61 | - |
| BGO_OH | **5** | 2.45 | - |

**Table 4** Thermochemistry of each reduction step at pH = 0 (see plot in Figure 6) and pH = 14 (see plot in Figure 7). For the different free energy terms see the text in Section 3.2. Energy and free energy variations in eV.

| | **ACIDIC CONDITIONS pH = 0** | | | |
|---|---|---|---|---|
| | **E** | **G** | **$G_w$** | **$G_{w/sol}$** |
| $BG + O_2 + 4H^+ + 4e^-$ | 4.46 | 4.64 | 4.64 | 4.64 |
| $BGOO + 4H^+ + 4e^-$ | 4.80 | 5.50 | 5.30 | 5.10 |
| $BGOOH + 3H^+ + 3e^-$ | 3.82 | 4.88 | 4.42 | 4.22 |
| $BGO + H_2O + 2H^+ + 2e^-$ | 1.81 | 2.19 | 2.13 | 2.04 |
| $BGOH + H_2O + H^+ + e^-$ | 0.75 | 1.46 | 1.09 | 0.94 |
| $BG + 2 H_2O$ | 0.00 | 0.00 | 0.00 | 0.00 |
| | **ALKALINE CONDITIONS pH = 14** | | | |
| | **E** | **G** | **$G_w$** | **$G_{w/sol}$** |
| $BG + O_2 + 2 H_2O + 4e^-$ | 1.15 | 1.33 | 1.33 | 1.33 |
| $BGOO + 2 H_2O + 4e^-$ | 1.49 | 2.19 | 1.99 | 1.78 |
| $BGOOH + OH^- + H_2O + 3e^-$ | 1.34 | 2.40 | 1.94 | 1.73 |
| $BGO + 2 OH^- + H_2O + 2e^-$ | 0.15 | 0.53 | 0.47 | 0.38 |
| $BGOH + 3 OH^- + 1e^-$ | -0.07 | 0.63 | 0.26 | 0.11 |
| $BG + 4 OH^-$ | 0.00 | 0.00 | 0.00 | 0.00 |

**Figures Captions**

**Figure 1-** Reaction paths for the first two reduction steps of the dissociative pathway. See Table 1.
**Figure 2-** Reaction paths for the first two reduction steps of the associative pathway. Values in parenthesis are obtained on the gas-phase optimized transition structure. See Table 2.
**Figure 3-** Reaction paths for the second two reduction steps for both the dissociative and associative pathways. See Tables 1 and 2.
**Figure 4-** Charged species along the associative pathway. See Tables 1 and 2.
**Figure 5-** Ball-and-stick representation of the intermediates of the ORR as catalyzed by B-doped graphene: top left, BGOO; top right, BGOOH; bottom left, BGO; bottom right, BGOH.
**Figure 6-** Free energy ($\Delta G_{w/sol}$) diagram for the ORR reaction catalyzed by B-doped graphene in acidic conditions (pH = 0), based on data in Table 4. Dashed line represents the optimal catalyst.
**Figure 7-** Free energy diagram ($\Delta G_{w/sol}$) for the ORR reaction catalyzed by B-doped graphene in alkaline conditions (pH = 14), based on data in Table 4. Dashed line represents the optimal catalyst.
**Figure 8-** Top panel: stability of the intermediates of ORR on B-doped graphene at pH = 0 at varying electrode potential. Bottom panel: surface Pourbaix diagram; the dash line represents the equilibrium potential at varying pH value.

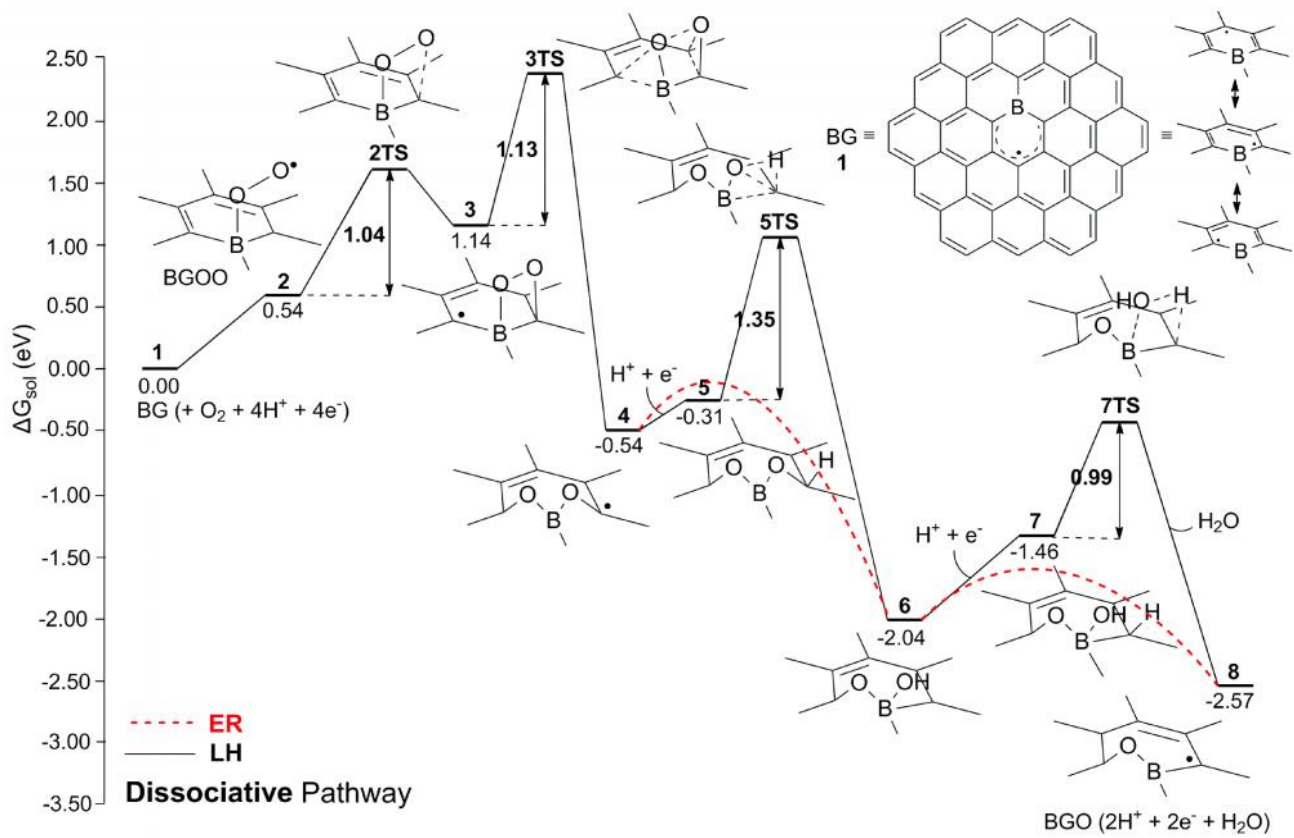

**Figure 1**

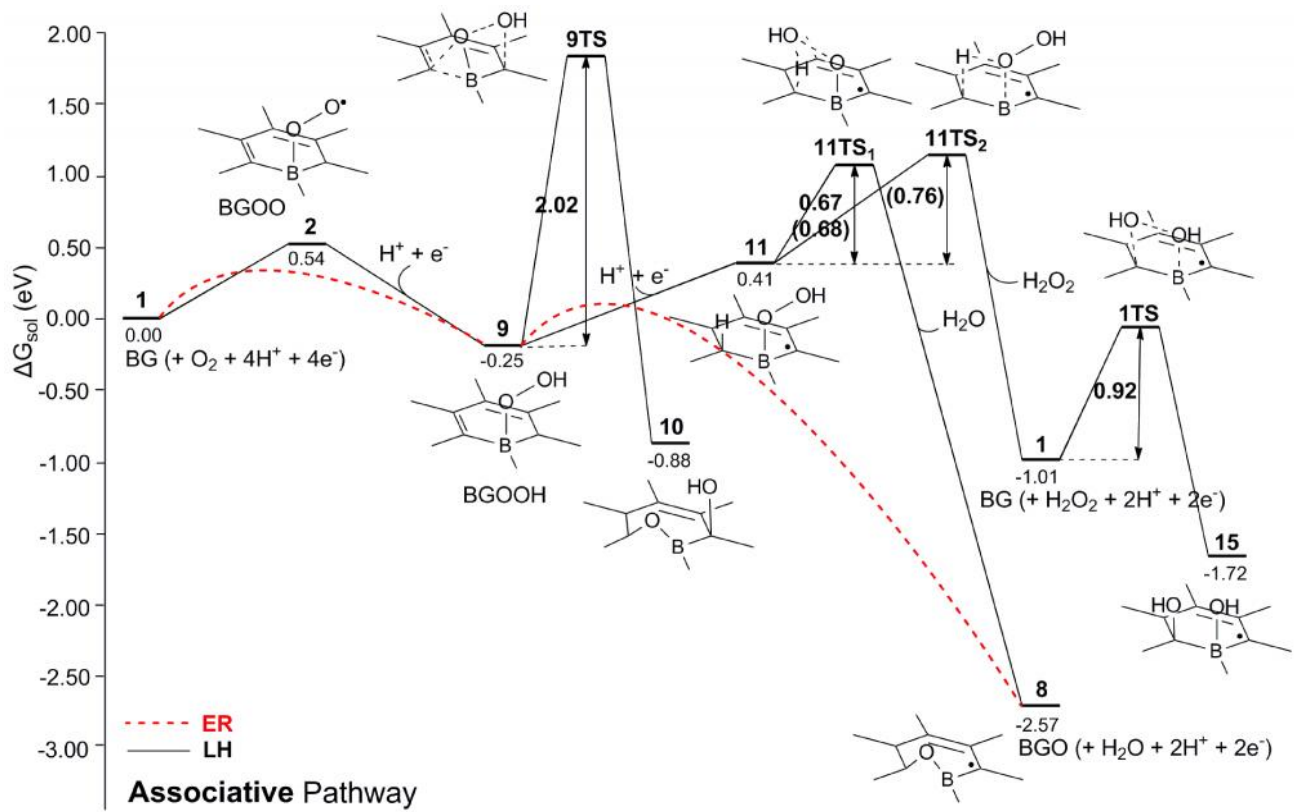

**Figure 2**

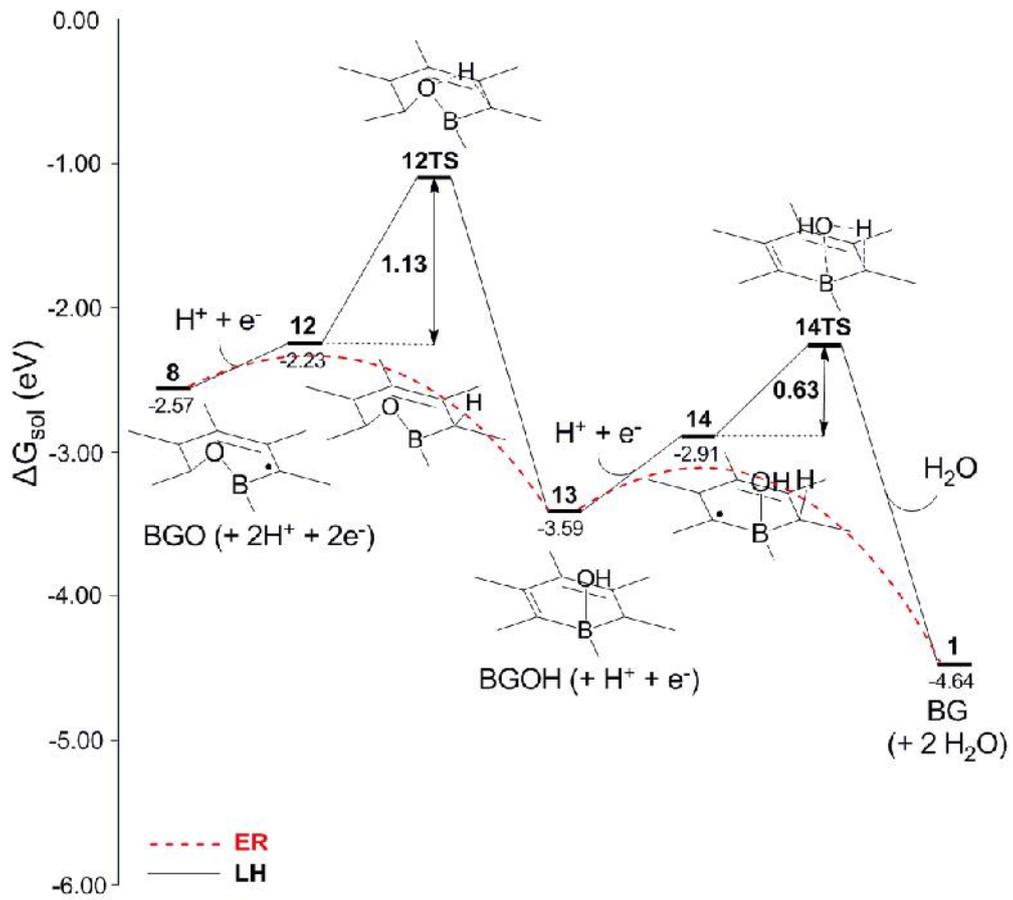

**Figure 3**

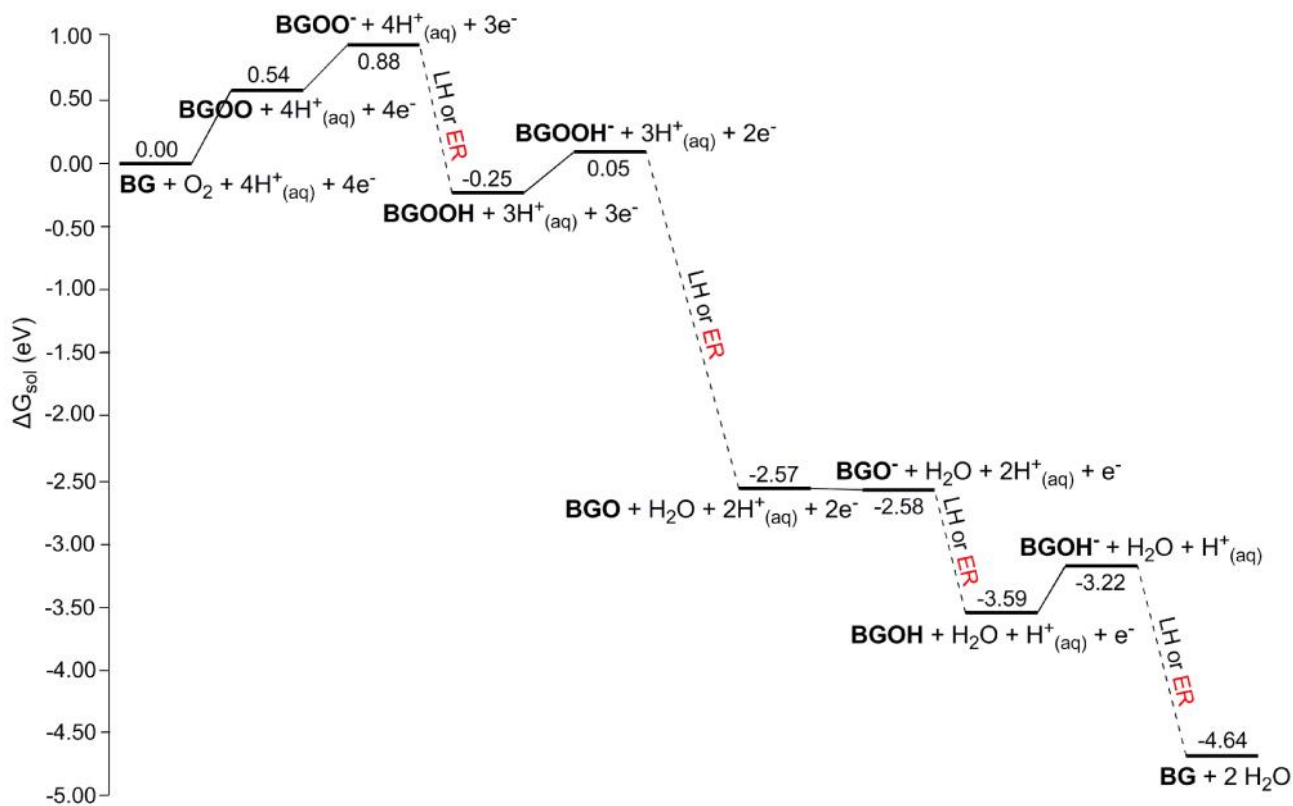

**Figure 4**

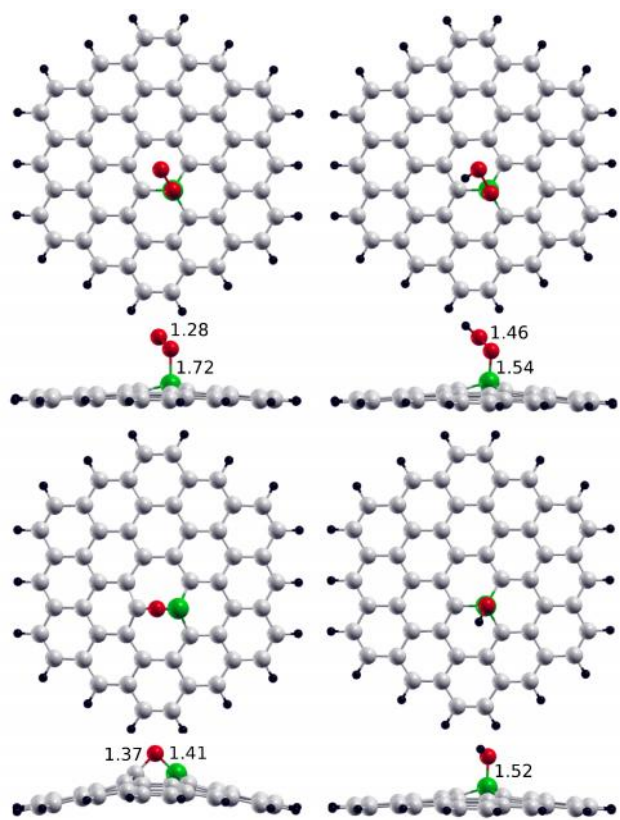

**Figure 5**

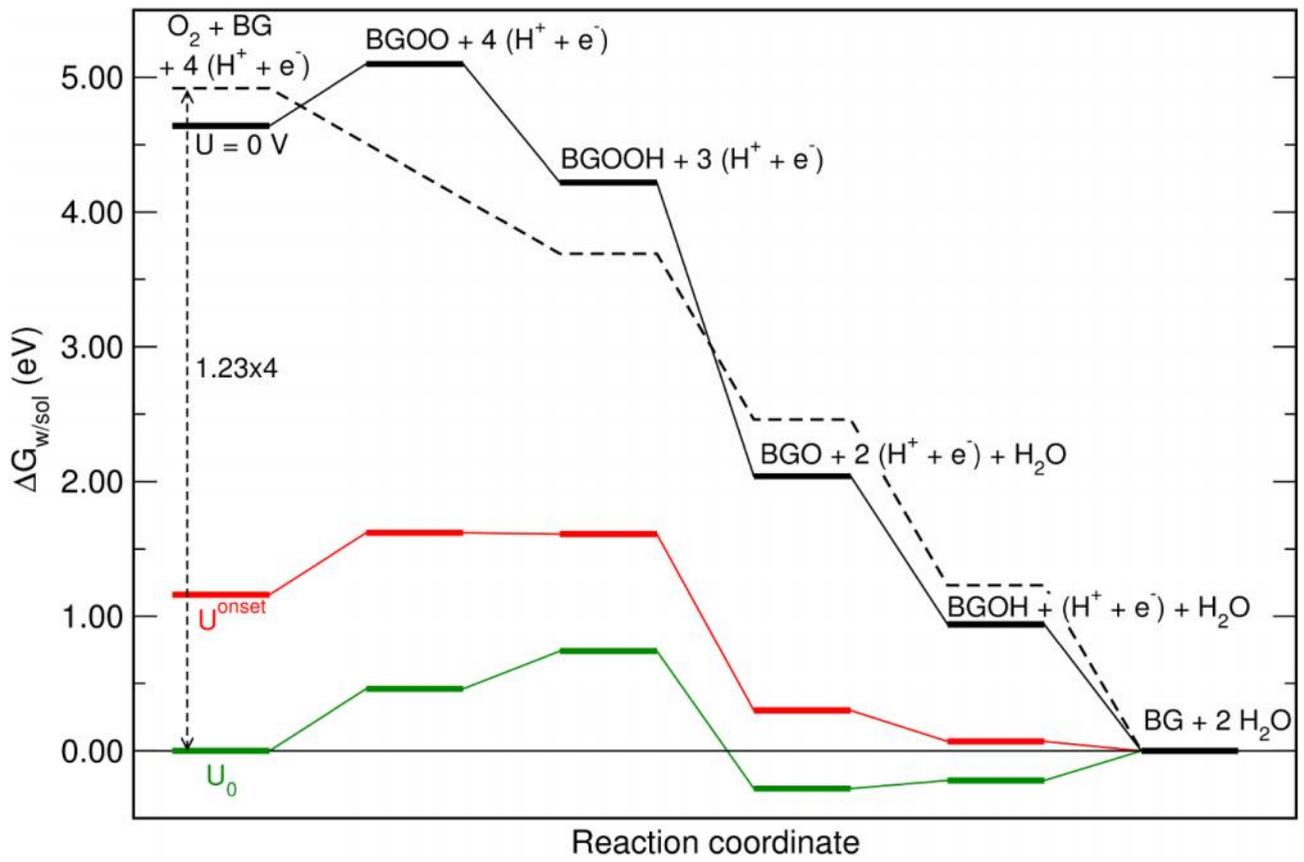

**Figure 6**

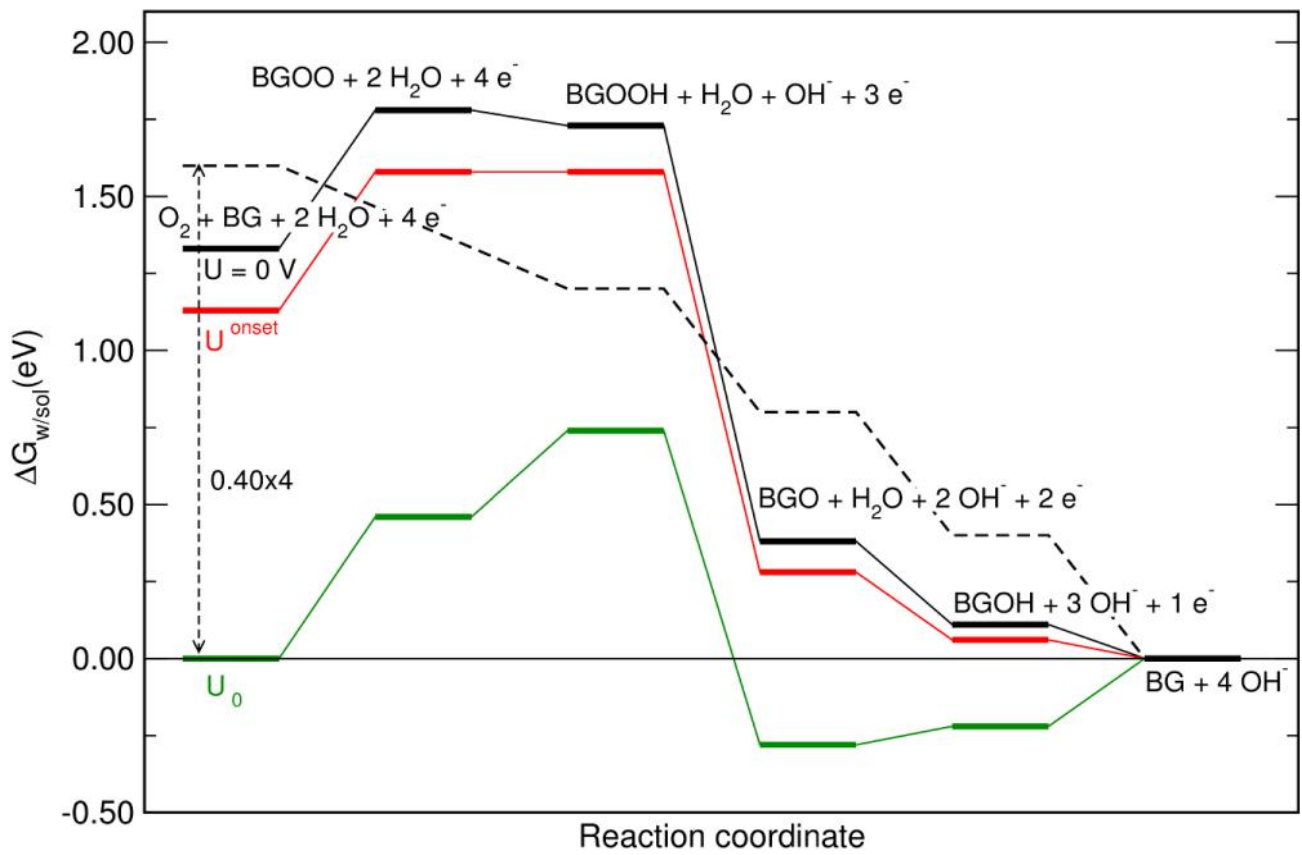

**Figure 7**

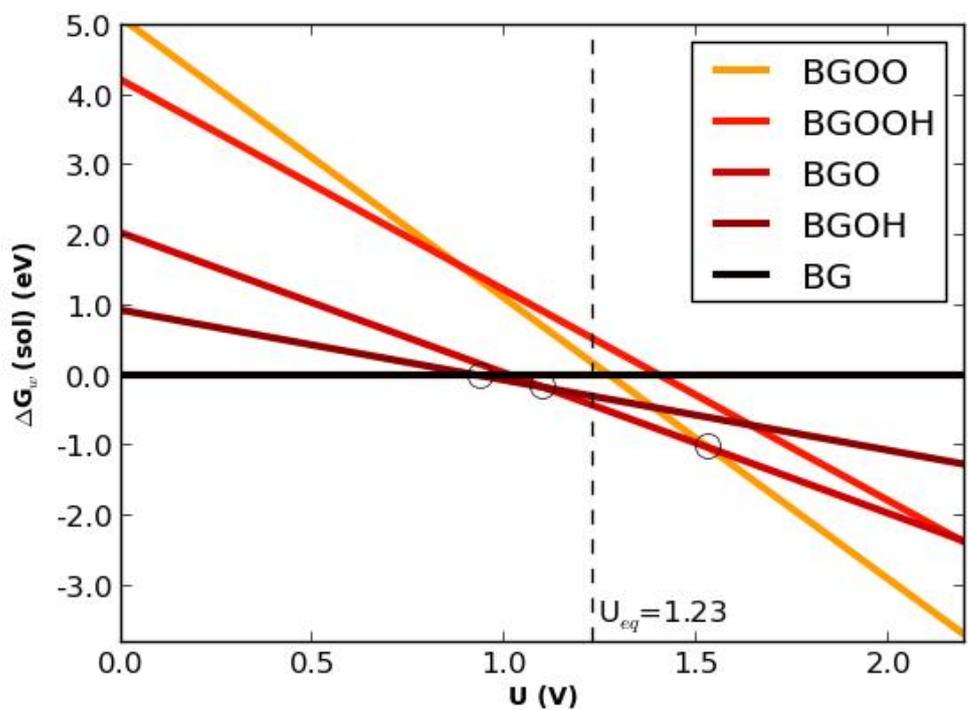
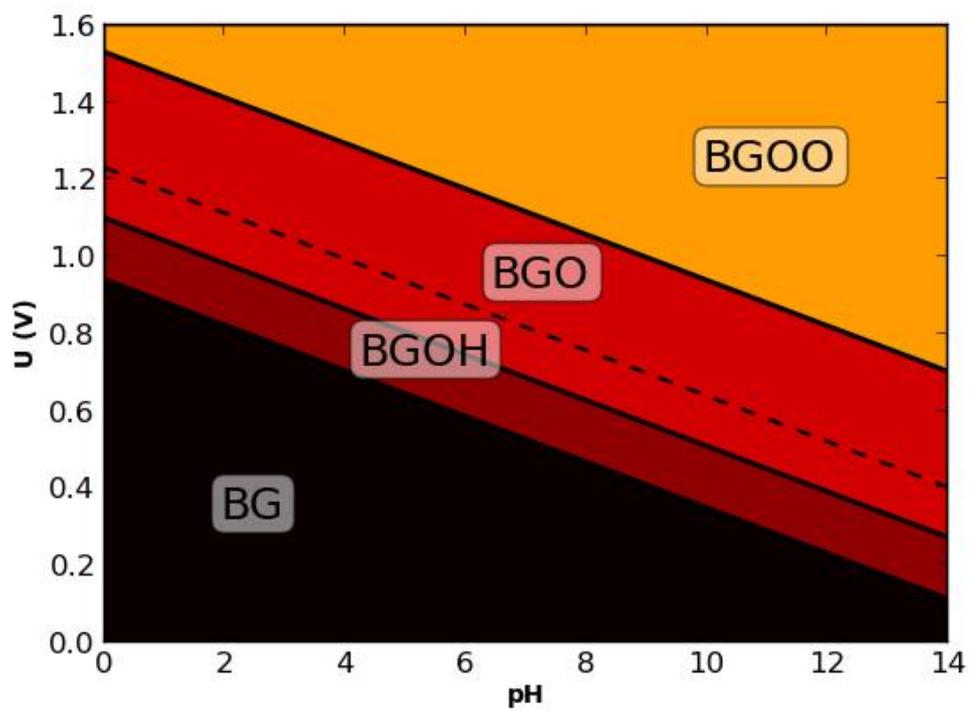

**Figure 8**

# References

[1] B. C. H. Steele, A. Heinzel, Nature 414 (2001) 345–352.
[2] K. Gong, F. Du, Z. Xia, M. Durstock, L. Dai, Science 323 (2009) 760–764.
[3] Y. Li, W. Zhou, H. Wang, L. Xie, Y. Liang, F. Wei, J.-C. Idrobo, S. J. Pennycook, H. Dai, Nat Nanotechnol. **7** (2012), 394–400.
[4] Z.-S. Wu , S. Yang , Y. Sun , K. Parvez , X. Feng , K. Müllen, J. Am. Chem. Soc. 134 (2012) 9082–9085.
[5] C. Choi, S. H. Park, L. Woo, ACS Nano 6 (2012) 7084–7091.
[6] Y. Li, Y. Zhao, H. Cheng, Y. Hu, G. Shi, L. Dai, L. Qu, J. Am. Chem. Soc. 134 (2012) 15–18.
[7] D. Geng, Y. Chen, Y. Chen, Y. Li, R. Li, X. Sun, S. Ye and S. Knights Energy Environ. Sci. 4 (2011) 760–764.
[8] L. Qu, Y. Liu, J.-B. Baek, L. Dai, ACS Nano 4 (2010) 1321–1326.
[9] Z. Yang, Z. Yao, G. Li, G. Fang, H. Nie, Z. Liu, X. Zhou, X. Chen, S. Huang, ACS Nano 6 (2012) 205–211.
[10] Z.-H. Sheng, H.-L. Gao, W.-J. Bao, F.-B. Wang, X.-H. Xia, J. Mater. Chem. 22 (2012) 390–395.
[11] L. Yang, S. Jiang, Y. Zhao, L. Zhu, S. Chen, X. Wang, Q. Wu, J. Ma, Y. Ma, Z. Hu, Angew. Chem. 123 (2011) 123 7270–7273.
[12] Y. Jiao, Y. Zheng, M. Jaroniec, S. Z. Qiao, J. Am. Chem. Soc. 136 (2014) 4394–4403.
[13] X. Fan, W. T. Zheng, J.-L. Kuo, RSC Adv. 3 (2013) 5498–5505.
[14] M. Kaukonen, A. V. Krasheninnikov, E. Kauppinen, R. M. Nieminen, ACS Catal. (3) 2013 159–165.
[15] F. Studt, Catal. Lett. 143 (2013) 58-60.
[16] L. Yu, X. Pan, X. Cao, P. Hu, X. Bao, J. Catal. 282 (2011) 183–190.
[17] A. W. Saidi, J. Phys. Chem. Lett. 4 (2013) 4160-4165.
[18] Y. Okamoto, App. Surf. Sci. 256 (2009), 335.
[19] L. Zhang, Z. Xia, J. Phys. Chem. C 115 (2011) 11170-11176.
[20] J. Zhang, Z. Wang, Z. Zhu, J. Mol. Model. 19 (2013) 5515-5521.
[21] H. Kim, K. Lee, S. I. Woo, Y. Jung, PCCP 13 (2011) 17505-17510.
[22] X. Bao, X. Nie, D. von Deak, E. J. Bidding, W. Luo, A. Asthagirl, U. S. Ozkan, C. M. Hadad, Top. Catal. 56 (2013) 1623-1633.
[23] L. Ferrighi, M. Datteo, C. Di Valentin, J. Phys. Chem. 118 (2014) 223–230.
[24] J. K. Nørskov, J. Rossmeisl, A. Logadottir, L. Lindqvist, J. R. Kitchin, T. Blgaard, H. Jonsson, J. Phys. Chem. B 108 (2004) 17886-17892.
[25] H. A. Hansen, J. Rossmeisl, J. K. Nørskov, PCCP 10 (2008) 3722-3730.
[26] Gaussian 09, Revision D.01, M. J. Frisch, G. W. Trucks, H. B. Schlegel, G. E. Scuseria, M. A. Robb, J. R. Cheeseman, G. Scalmani, V. Barone, B. Mennucci, G. A. Petersson, et al. Gaussian, Inc. (2009), Wallingford CT.
[27] A. D. Becke, J. Chem. Phys. 98 (1993) 5648-5652.
[28] C. Lee, W. Yang, R. G. Parr, Phys. Rev. B 37 (1988) 785-789.
[29] J. Tomasi, B. Mennucci, R. Cammi, Chem. Rev. 105 (2005) 2999-3093.
[30] A. V. Marenich, C. J. Cramer, D. G. Truhlar, J. Phys. Chem. B 113 (2009) 6378-96.
[31] S. Grimme, J. Antony, S. Ehrlich. H. Krieg, J. Chem. Phys.132 (2010) 154104.
[32] S. Grimme, S. Ehrlich, L. Goerigk, J. Comp. Chem. 32 (2011) 1456-65.
[33] C.-G. Zhan, D. A. Dixon, J. Phys. Chem. A 105 (2001) 11534-11540
[34] M. D. Tissandier, K. A. Cowen, W. Y. Feng, E. Gundlach, M. H. Cohen, A. D. Earhart, J. V. Coe, T. R. Tuttle, Jr., J. Phys. Chem. A 102 (1998) 7787-7794.
[35] J. A. Keith, G. Jerkiewicz, T. Jacob, ChemPhysChem 11 (2010) 2779-2794.

# BORON-DOPED GRAPHENE AS ACTIVE ELECTROCATALYST FOR OXYGEN REDUCTION REACTION AT A FUEL-CELL CATHODE

Gianluca Fazio, Lara Ferrighi, Cristiana Di Valentin[*]

*Dipartimento di Scienza dei Materiali, Università di Milano-Bicocca*

*via Cozzi 55 20125 Milano Italy*

## Supplementary Content

**Table S1** Electronic energies and free energies for all intermediates and transition structures in Figure 1 and 3.

| Dissociative Pathway – Table of Intermediates | | | | | | | | |
|---|---|---|---|---|---|---|---|---|
| NAME | Intermediate | E(Ha) | G(Ha) | $O_2$[*] | ½ $H_2$[*] | $H_2O$[*] | ΔG(eV) | ΔG[‡](eV) |
| 1 | BG | -2055.66193 | -2055.18565 | 1 | 4 | 0 | 0.00 | |
| 1TS | TS | -2206.01963 | -2205.54060 | 0 | 4 | 0 | - | |
| 2 | $O_{2(ads)}$ | -2206.01969 | -2205.54060 | 0 | 4 | 0 | 0.86 | |
| 3TS | TS(O-O) | -2205.98725 | -2205.50659 | 0 | 4 | 0 | 1.79 | 0.93 |
| 3 | $O-O_{(ads)}$ | -2205.99124 | -2205.50957 | 0 | 4 | 0 | 1.71 | |
| 3TS | $TS(O_{(ads)}+O_{(ads)})$ | -2205.96055 | -2205.46747 | 0 | 4 | 0 | 2.85 | 1.15 |
| 4 | $O_{(ads)}+O_{(ads)}$ | -2206.06784 | -2205.58628 | 0 | 4 | 0 | -0.38 | |
| 5 | $O_{(ads)}+ O_{(ads)}+ H_{(ads)}$ | -2206.66261 | -2206.16664 | 0 | 3 | 0 | -0.14 | |
| 5a | H in meta | -2206.64736 | | | | | | |
| 5b | H in para | -2206.66005 | | | | | | |
| 5TS | TS | -2206.59937 | -2206.10783 | 0 | 3 | 0 | 1.46 | 1.60 |
| 6 | $O_{(ads)}+ OH_{(ads)}$ | -2206.71175 | -2206.21862 | 0 | 3 | 0 | -1.56 | |
| 7 | $O_{(ads)}+ OH_{(ads)}+ H_{(ads)}$ | -2207.28605 | -2206.78346 | 0 | 2 | 0 | -0.90 | |
| 7TS | TS | -2207.25773 | -2206.75897 | 0 | 2 | 0 | -0.23 | 0.67 |
| 8 | $O_{(ads)}$ | -2130.86274 | -2130.38371 | 0 | 2 | 1 | -2.45 | |
| | $O_{2(g)}$ | | -150.386661 | | | | | |
| | $H_{2(g)}$ | | -1.178103 | | | | | |
| | $H_2O_{(aq)}$ | | -76.456680 | | | | | |

* number of molecules considered in the ΔG calculation.

[*] Corresponding author: cristiana.divalentin@mater.unimib.it, +390264485235.

**Table S2** Electronic energies and free energies for all intermediates and transition structures in Figure 2 and 3.

| \<colspan=8\> Associative Pathway – Table of Intermediates | | | | | | | | |
|---|---|---|---|---|---|---|---|---|
| NAME | Intermediate | E(Ha) | G(Ha) | $O_2^*$ | $½ H_2^*$ | $H_2O^*$ | $H_2O_2^*$ | ΔG(eV) | ΔG$^‡$(eV) |
| 1 | BG | -2055.66193 | -2055.18565 | 1 | 4 | 0 | 0 | 0.00 | |
| 2TS | TS(BGOO) | -2206.01963 | -2205.54060 | | | | | - | |
| 2 | BGOO | -2206.01969 | -2205.54060 | 0 | 4 | 0 | 0 | 0.86 | |
| 9 | BGOOH | -2206.64403 | -2206.15235 | 0 | 3 | 0 | 0 | 0.25 | |
| 9a | (in ortho) | -2206.61790 | | | | | | | |
| 9b | (in para) | -2206.61851 | | | | | | | |
| 9TS | TS(O$_{(ads)}$ + OH$_{(ads)}$) | -2206.57383 | -2206.08173 | 0 | 3 | 0 | 0 | 2.17 | 1.92 |
| 10 | O$_{(ads)}$ + OH$_{(ads)}$ | -2206.67841 | -2206.18368 | 0 | 3 | 0 | 0 | -0.61 | |
| 11 | OOH$_{(ads)}$ + H$_{(ads)}$ | -2207.21614 | -2206.71591 | 0 | 2 | 0 | 0 | 0.94 | |
| 11a | H in para | -2207.19939 | | | | | | | |
| 11TS$_2$ | TS(H$_2$O$_2$) | -2207.19002 | -2206.69520 | 0 | 2 | 0 | 0 | 1.50 | 0.56 |
| 1 | BG + H$_2$O$_2$ | -2055.66193 | -2055.18565 | 0 | 2 | 0 | 1 | -1.01 | |
| 1TS | TS(OH$_{(ads)}$ + OH$_{(ads)}$) | -2207.23744 | -2206.73739 | 0 | 2 | 0 | 0 | 0.35 | 1.36 |
| 11TS$_1$ | TS(H$_2$O) | -2207.19133 | -2206.69367 | 0 | 2 | 0 | 0 | 1.54 | 0.61 |
| 15 | OH$_{(ads)}$ + OH$_{(ads)}$ | -2207.29689 | -2206.79305 | 0 | 2 | 0 | 0 | -1.16 | |
| 8 | BGO | -2130.86274 | -2130.38371 | 0 | 2 | 1 | 0 | -2.45 | |
| 12 | O$_{(ads)}$ + H$_{(ads)}$ | -2131.45278 | -2130.96123 | 0 | 1 | 1 | 0 | -2.13 | |
| 12a | H on B | -2131.43002 | | | | | | | |
| 12b | H on ether C | -2131.45094 | | | | | | | |
| 12c | H in meta1 | -2131.43262 | | | | | | | |
| 12d | H in meta2 | -2131.42815 | | | | | | | |
| 12e | H in para | -2131.45019 | | | | | | | |
| 12cTS | TS para-meta1 diffusion | -2131.40188 | | | | | | | |
| 12TS | TS(BGOH) | -2131.40541 | -2130.91860 | 0 | 1 | 1 | 0 | -0.97 | 1.16 |
| 12TSb | from H in ether | -2131.38133 | | | | | | | |
| 12TSe | from H in para | -2131.37919 | | | | | | | |
| 13 | BGOH | -2131.48970 | -2130.99979 | 0 | 1 | 1 | 0 | -3.18 | |
| 13a | BGOH bound to C(ortho) | -2131.46661 | -2130.97544 | | | | | | |
| 14 | OH$_{(ads)}$ + H$_{(ads)}$ | -2132.06263 | -2131.56308 | 0 | 0 | 1 | 0 | -2.48 | |
| 14a | OH on B e H in para | -2132.04471 | | | | | | | |
| 14TS | TS(H$_2$O) | -2132.04270 | -2131.54763 | 0 | 0 | 1 | 0 | -2.06 | 0.42 |
| 1 | BG | -2055.66193 | -2055.18565 | 0 | 0 | 2 | 0 | -4.64 | |
| | H$_2$O$_{2(aq)}$ | | -151.601897 | | | | | | |

* number of molecules considered in the ΔG calculation.

**Figure S1** Spin density plot for BGOO (**2**) species.

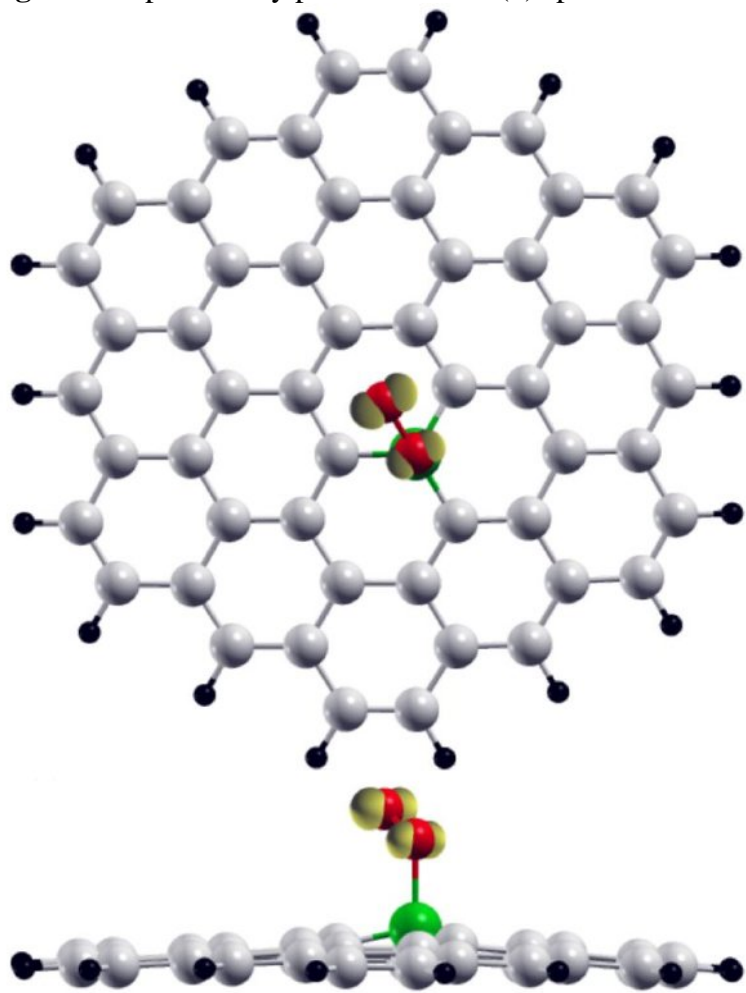

**Figure S2** Comparison of the ORR reaction profile on BG with that on Pt as computed in the reference: J. A. Keith, G. Jerkiewicz, T. Jacob, ChemPhysChem 11 (2010) 2779-2794.

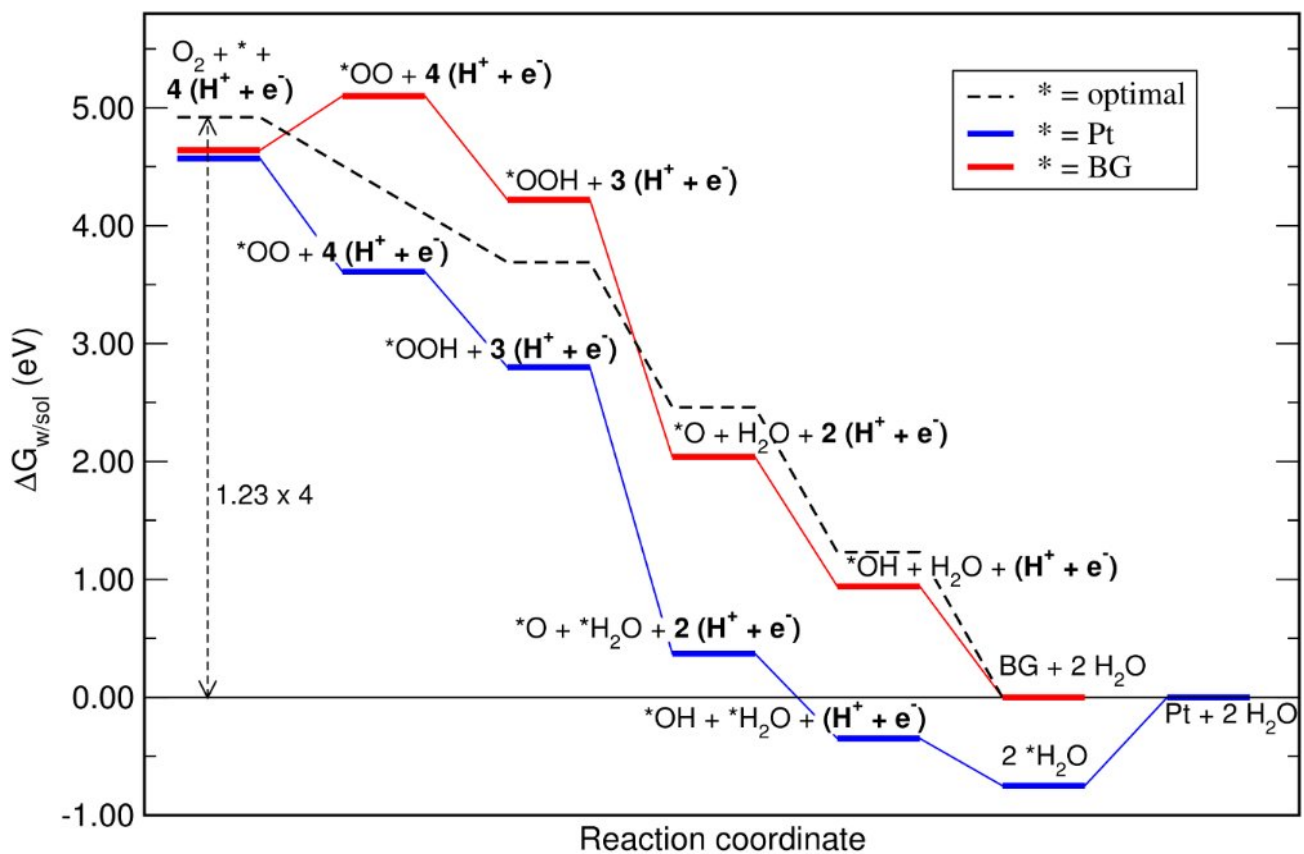